\documentclass[floatfix,twocolumn,showpacs,preprintnumbers,amsmath,amssymb]{revtex4}

\usepackage{amsfonts}

\begin{document}

\title{A Problem in Paleobiology}
\author{Barry D.\ Hughes}
\address{Department of Mathematics and Statistics,
University of Melbourne,
Victoria 3010, Australia}
\author{William J.\ Reed}
\address{Department of Mathematics and Statistics,
University of Victoria,
Victoria, British Columbia, Canada V8W 3P4}
\date{20 November 2002}

\begin{abstract}
We present a stochastic model for the size of a taxon in paleobiology,
in which we allow for the evolution of new taxon members, and both
individual and catastrophic extinction events.  The model uses ideas from 
the theory of birth and death processes. Some general properties
of the model are developed, and a fuller discussion is given for specific
distributions of the time between catastrophic extinction events.
Long tails in the taxon size distribution arise naturally from the model.    
\end{abstract}%

\maketitle

Random processes leading to probability distributions with slowly-decaying algebraic tails have been of
considerable recent interest in physics, although they have been sporadically studied in other areas for 
many years. As probability laws with $\Pr\{X>x\}\propto x^{-\alpha}$ as $x\to\infty$ are naturally associated
with scaling properties, self-similarity and fractals, it is tempting to propose fractal underlying
mechanisms to explain the occurrence of these laws in particular contexts in nature.  

We address a problem of interest in paleobiology, where long-tailed distributions arise, and propose a
model that requires no explicitly fractal underlying mechanism to explain these distributions. 
The model predicts the distribution of the number of elements in a taxon,
for example the number of species in a genus, and our analysis covers both the distribution of species
currently in existence, and the distribution of all species in the genus that have ever existed, some of
which may now be extinct.

Several other problems in taxonomy and genetics involve similar mathematical analysis, 
and the authors have briefly addressed elsewhere \cite{ReedHughesLiveTaxa,ReedHughesDCDSB} the fitting of
models of this type to real biological data. More generally, the basic modelling approach in this paper
involves killing at an exponentially distributed time a stochastic process for which the mean grows
exponentially in time.  The appearance of power-law distributions in such contexts has been discussed in
general by the authors elsewhere \cite{ReedHughesPREBrief}, and has applications in social phenomena
\cite{ReedHughesPhysicaA} and other contexts not covered in the present paper.

Modelling of taxon size has been of sporadic interest in the literature for some time
\cite{yule,raup,chu}, and there has been a steady accumulation of relevant data on both extinct and surviving 
taxonomic groups \cite {burlando1,burlando2}. We shall not address mechanisms for species interaction in an ecosystem that may drive extinction or species
proliferation, nor shall we address the shape of evolutionary trees (cladograms).

In our model, a taxon comes into existence with a single representative species, genus or family at time
$t=0$.  The number of members of the taxon grows with time  as species mutate.  Some species produce many new
species, but any species  may also become extinct, and a natural model for the proliferation and extinction
of taxon members is the linear birth--death process (see, for example, \cite{coxmiller}, pp.~165--167 and
pp.~265--266).  A taxon member (referred to hereafter as a species) has in the time interval $(t,t+h)$ a
probability $\lambda h+o(h)$ of `giving birth' to a new species, and a probability $\mu h+o(h)$ of `dying'.
Assuming independence of speciation and extinction events, if there are $n$ species present at time $t$, the
probability of one speciation occurring in $(t,t+h)$ is $\lambda nh+o(h)$, and the probability of one
extinction is $\mu nh+o(h)$.
Let $M_t$ denote the number of species that have ever existed (whether currently alive or not) at time $t$
and $N_t$ the number of species currently alive at time $t$, and write
$p_{m,n}(t)=\Pr\{M_t=m,\, N_t=n\}$.
Then for $h>0$,
\begin{eqnarray*}
p_{m,n}(t+h)\kern-4pt&=&\kern-4pt
[1-n(\lambda+\mu)h+o(h)]p_{m,n}(t)\\ 
&&+[\lambda h+o(h)](n-1)p_{m-1,n-1}(t)\\ 
&&+[\mu h+o(h)](n+1)p_{m,n+1}(t)+o(h). 
\end{eqnarray*}
The first term on the right-hand side corresponds to no change in $M_t$ or $N_t$
in the time interval $(t,t+h)$, the second to one birth, and the third to one death.
All other events have probability $o(h)$.  Subtracting $p_{m,n}(t)$ from both sides,
dividing by $h$ and letting $h\to 0$, we deduce the differential-difference equation
\begin{eqnarray*}
{d\over{dt}}p_{m,n}(t)&=&-(\lambda+\mu)np_{m,n}(t)+\lambda(n-1)p_{m-1,n-1}(t)\\ 
&&+\mu(n+1)p_{m,n+1}(t).
\end{eqnarray*}
This equation is valid for all integer $m$ and $n$ provided that we adopt the convention that
$p_{m,n}(t)=0$ if $m\leq 0$ or $n<0$.
We measure time from the appearance of the first individual, so that
$p_{m,n}(0)=1$ if $m=n=1$; $p_{m,n}(0)=0$ otherwise.
Introducing the generating function
$P(\xi,\zeta,t)=\mathbb{E}\{\xi^{M_t}\zeta^{N_t}\}$ (where $\mathbb{E}$ denotes expectation) 
one may readily derive \cite{kendall} the partial differential equation
$P_t=\left\{\xi\zeta^2\lambda-(\lambda+\mu)\zeta+\mu\right\}P_\zeta$
with initial condition
$P(\xi,\zeta,0)=\xi\zeta$. 
 Using the method of characteristics  one finds \cite{kendall} that
\begin{equation}
P(\xi,\zeta,t)=\xi\,
{{z_2(\zeta -z_1)e^{\lambda\xi z_1t}
+z_1(z_2-\zeta)e^{\lambda\xi z_2t}}
\over{(\zeta -z_1)e^{\lambda\xi z_1t}
+(z_2-\zeta)e^{\lambda\xi z_2t}}},\label{eKendallSolution}
\end{equation}
where $z_1$ and $z_2$ denote the roots of the quadratic equation
$\xi\lambda z^2-(\lambda+\mu)z+\mu=0$. 
Since the probability that $M_t$ species have ever existed up to time $t$
is given by $\Pr\{M_t=m\}=\sum_{n=1}^\infty p_{mn}$, setting $\zeta=1$ in
Eq.\ (\ref{eKendallSolution}) we recover the generating function for the probability that $M_t$
species have ever existed up to time $t$:
\[ 
\mathbb{E}\{\xi^{M_t}\}={{x_2(\lambda\xi -x_1)e^{x_1t}
+x_1(x_2-\lambda\xi)e^{x_2t}}
\over{\lambda[(\lambda\xi -x_1)e^{x_1t}
+(x_2-\lambda\xi)e^{x_2t}]}}. 
\] 
For brevity we have let $x_i=\lambda\xi z_i$, so that $x_1$ and $x_2$ 
are the roots of $x^2-(\lambda+\mu)x+\lambda\mu\xi=0$.
These roots are distinct for all $\mu$ and $\lambda$ when $\xi<1$.  
The roots become $\lambda$ and $\mu$ when $\xi=1$, so that the case $\lambda=\mu$ is
degenerate for $\xi=1$, but this presents no difficulties in the subsequent analysis.
Using the results that 
$x_1+x_2=\lambda+\mu$ and $x_1x_2=\lambda\mu\xi$, establishes
the form of the generating function needed below:
\begin{equation}
\mathbb{E}\{\xi^{M_t}\}=\xi+{{\lambda\xi(\xi-1)[e^{(x_2t} - e^{x_1t}]}
\over{(x_2-\lambda\xi)e^{x_2t}-(x_1 - \lambda\xi)e^{x_1t}}}.\label{eNiceforM}
\end{equation}

To make contact with better known results on the number of species currently alive, we note that setting
$\xi=1$ gives $z_1=1$ and $z_2=\mu/\lambda$ and 
Eq.\ (\ref{eKendallSolution}) becomes \cite{cfCoxMandGrim} 
\[ 
\mathbb{E}\{\zeta^{N_t}\}=\left\{\begin{array}{ll}
\displaystyle {{\mu(1-\zeta)-(\mu-\lambda\zeta)e^{-t(\lambda-\mu)}}
\over{\lambda(1-\zeta)-(\mu-\lambda\zeta)e^{-t(\lambda-\mu)}}}
&\mbox{if}~\lambda\neq\mu,\\ \\
1-(1-\zeta)/[1+\lambda t(1-\zeta)]^{-1}
&\mbox{if}~\lambda=\mu.
\end{array}\right.
\] 
That $\langle N_t\rangle=e^{(\lambda-\mu)t}$ 
follows by differentiation, while expansion of the generating function gives (\cite{coxmiller}, p.~166) 
$\Pr\{N_t=0\}=(\mu-\mu e^{-t(\lambda-\mu)})/(\lambda-\mu e^{-t(\lambda-\mu)})$;
for $n\geq 1$,
\[
\Pr\{N_t=n\}={{(\lambda-\mu)^2e^{-t(\lambda-\mu)}}\over{[\lambda-\mu}e^{-t(\lambda-\mu)}]^2}\,
\left\{{{\lambda-\lambda e^{-t(\lambda-\mu)}}\over{\lambda-\mu e^{-t(\lambda-\mu)}}}\right\}^{n-1}.
\]
In the limiting case $\lambda=\mu$,  
$\Pr\{N_t=0\}=\lambda t/(1+\lambda t)$ and
$\Pr\{N_t=n\}=(\lambda t)^{n-1}/(1+\lambda t)^{n+1}$ for $n\geq 1$.
When $\mu=0$ (that is, there is a pure birth process) the solution reduces to that
found by Yule \cite{yule} in his model of species evolution under a speciation rate $\lambda$.

There is considerable evidence for major catastrophic extinctions occurring within
a relatively short period, these extinctions having been attributed to various causes,
including major meteorite impacts \cite{alvarez} and a hypothesised purely biotic mechanism called 
coevolutionary avalanches \cite{newman}. To include catastrophic
extinctions in our model, we require the probability density function $f(t)$ for 
the time $T$ between the start of a taxon and the next catastrophe.  In the
analysis below we carry a general $f(t)$ as far 
as possible.  The three specific models discussed here are proposed with
some diffidence, though each has a certain natural appeal, and each may apply to
appropriate subsets of paleological data.  The common thread to all three models is that as
$t\to\infty$,
\begin{equation}
f(t)\sim\hbox{constant}\times t^q e^{-\theta t},\label{eCatastropy1}
\end{equation}
with $1/\theta$ the mean time between catastrophic extinction events and
either $q=0$ or $q=-1$. The small-$t$ behavior is different in the three models, but
this difference does not affect the dominant asymptotic behavior of the taxon size distribution. 
Using Eq.\ (\ref{eCatastropy1}) with some flexibility as to the value of $q$ seems 
a reasonable approach.  

(a) {\em The pure exponential model.\/} As a first model one may assume that
$f(t)=\theta e^{-\theta t}$ for $t>0$.
This asserts that the waiting time for the next catastrophe is
exponentially distributed, but effectively considers only one taxon:
no account is taken of the fact that in a long time interval, 
many taxons should be initiated, while in a short time interval, it is
likely that no taxons will be initiated.  Subtle conditional probability
effects are ignored.

For  models (b) and (c) below, we assume that catastrophic extinction events occur in a Poisson process at
rate $\theta$, while taxon initiations occur in a Poisson process at rate $\rho$. Thus the probability
density function for the waiting time between extinctions is 
$\psi(t)=\theta e^{-\theta t}$, $t>0$,
while the waiting-time density for the start of the next taxon is 
$\chi(t)=\rho e^{-\rho t}$, $t>0$. 

(b) {\em The first new taxon model.\/}
Consider the time to the next catastrope for the {\em first\/} taxon initiated after the previous
catastrophe. If we condition on the time $T$ between catastrophes, 
the conditional waiting-time density for appearance
of a taxon is
$\rho e^{-\rho t}/(1-e^{-\rho T})$, $0<t<T$.
The time between the appearance of the taxon and the next catastrophe therefore has the 
probability density function
$\eta(t\,|\,T)=\rho e^{\rho (t-T)}/(1-e^{-\rho T})$,  $0<t<T$.
We now average over $T$ to deduce for the time from taxon commencement to the next catastrophe
the density
\[
f(t)=\int_0^\infty \theta e^{-\theta \tau}\eta(t\,|\,\tau)d\tau
=\rho\theta e^{\rho t}\int_t^\infty 
{{e^{-\rho \tau-\theta\tau}d\tau}\over{1-e^{-\rho \tau}}}.
\] 
It can be shown that
$f(t)\sim \theta\ln[1/(\rho t)]$ as $t\to 0$, while
$f(t)\sim [\rho\theta/(\rho+\theta)]\,e^{-\theta t}$ as $t\to\infty$.

(c) {\em Uniform taxon nucleation between catastrophes.\/}
The probability that there is at least one taxon initiated in the time
interval of duration $\tau$ between two successive catastrophes is $1-e^{-\rho \tau}$.
It is known \cite{snyder} that for a Poisson process with rate $\rho$, conditional on 
there being $n$ occurrences in a time interval of length $\tau$, the occurrence times have the same 
distribution as the order statistics of a set of $n$ independent times, each uniformly distributed
on the interval of length $\tau$.  This suggests as a model for the 
probability density function for the time to the next catastrophe
\[
f(t)={{\rho+\theta}\over\rho}\int_t^\infty 
{{[1- e^{-\rho\tau}]}\over\tau}\,\theta e^{-\theta\tau}d\tau.
\]
The prefactor $(\rho+\theta)/\rho$ is inserted to ensure that $f(t)$ is non-defective, that is, 
integrates to unity.  
As $t\to\infty$,
\[
f(t)={{(\rho+\theta)e^{-\theta t}}\over{\rho t}}\bigl[1+O(t^{-1})\bigr]\,
\bigl[1+O(e^{-\rho t})\bigr].
\]

{\em The size of a surviving taxon.\/} 
We address briefly the distribution of the number $N$ of species in a taxon that are living
just before a catastrophic extinction event occurs; equivalently this is asking for the distribution of taxon size
today, the detail residing in the probability density function $f(t)$ for the time
since the taxon began.   The case $\lambda<\mu$, in which a taxon is driven rapidly to extinction, 
is not considered. We shall consider only the case
$f(t)=\theta e^{-\theta t}$, $t\geq 0$.
Since relatively simple expressions 
for $\Pr\{N_t=n\}$ are available, the direct calculation of the distribution of
\[
\Pr\{N=n\}=\int_0^\infty \Pr\{N_t=n\} f(t)dt 
\]
becomes possible; the details are equivalent to those 
in a model of live taxa where both species and genera proliferate \cite{ReedHughesLiveTaxa}
and will not be given here.

In the case $\lambda=\mu$, the distribution of $N$ is reasonably rapidly decaying, though its 
dominant form is subtle:
\[
\Pr\{N=n\}\sim \pi^{1/2}(\theta/\lambda)^{5/4}n^{-3/4}e^{-2(\theta/\lambda)^{1/2}n^{1/2}}.
\]
The stretched exponential behavior is typical of the crossover behavior in problems of stochastic
processes or statistical physics when exponential decay  degenerates to algebraic decay as a parameter (here
$\mu$) passes through a critical value (here $\lambda$).

If $\lambda>\mu$, we find  
$\Pr\{N=n\}\sim \hbox{constant}\times n^{-1-\theta/(\lambda-\mu)}$, 
so that $\Pr\{N\geq n\}\sim\hbox{constant}\times n^{-\theta/(\lambda-\mu)}$ as $n\to\infty$. 
The mean taxon size is infinite if $\lambda\geq\mu+\theta$.
Applications of these formulae to real data are given elsewhere \cite{OurMLestimates}.

{\em Proliferation between catastrophes.\/} 
The problem of greater palaeobiological interest
concerns the number of species that ever belong to a taxon.
As before, let $f(t)$ be the waiting-time density for the time
$T$ after the emergence of a  taxon to the next global extinction.  
The case $f(t)=\theta e^{-\theta t}$
is of most interest, but we carry generality when we may.
Let the random variable $M$  be the number 
of species in a taxon that exists only between two successive catastrophes.
With $M_t$ the number of species that have ever existed up to time $t$,  we have
\[
p_m=\Pr\{M=m\}=\int_0^\infty \Pr\{M_t=m\}f(t)dt.
\]
Using Eq.\ (\ref{eNiceforM}), the generating function for $p_m$ is given by
\begin{equation}
\phi(\xi)= \sum_{m=1}^\infty p_m\xi^m=1+(\xi-1)\{1+\xi\chi(\xi)\},\label{eB7a}
\end{equation}
where
\begin{equation}
\chi(\xi)=\int_0^\infty {{\lambda[e^{x_2t}-e^{x_1t}]f(t)dt}
\over{(x_2-\lambda\xi)e^{x_2t} - (x_1-\lambda\xi)e^{x_1t}}}.\label{eB7b}
\end{equation}
We need to determine the asymptotic behavior 
of $\chi(\xi)$ near $\xi=1$.  
If the function $\chi(\xi)$ and its first derivative $\chi\,'(\xi)$, respectively, remain finite at
$\xi=1$, then the expected value $\langle M\rangle$ of $M$ and the variance $\hbox{Var}\{M\}$ of $M$ 
are finite, and we have 
$\langle M\rangle=1+\chi(1)$ and $\hbox{Var}\{M\}=2\chi\,'(1)+\chi(1)-\chi(1)^2$.
The function $\chi(\xi)$ is symmetric under interchange of $x_1$ and $x_2$.
We shall identify $x_2$ with the root that approaches $\lambda$ as $\xi\to 1$,
and $x_1$ with the root that approaches $\mu$  as $\xi\to 1$.
Solving the quadratic equation for $x_1$ and $x_2$ exactly and expanding the solutions for
$1-\xi\to 0$, we record for later use that
\begin{eqnarray*}
x_1&=&\mu-\lambda\mu(1-\xi)/(\lambda-\mu)+O([1-\xi]^2),\\
x_2&=&\lambda+\lambda\mu(1-\xi)/(\lambda-\mu)+O([1-\xi]^2).
\end{eqnarray*}
Provided that the integral in Eq.\ (\ref{eB7b}) converges for $\xi=1$, we 
find that the expected value of $M$ is
\begin{equation}
\langle M\rangle
=1+\int_0^\infty {{\lambda[e^{(\lambda-\mu)t}-1]f(t)dt}\over{\lambda-\mu}}.\label{eB13}
\end{equation}
For $\lambda<\mu$, the mean is finite
for every density $f(t)$. 
The degenerate case $\lambda=\mu$ can be analysed 
separately, or by taking the limit $\lambda\to\mu$
from below inside the integral in Eq.\ (\ref{eB13}), 
giving $\langle M\rangle=1+\lambda\langle T\rangle$,
so $\langle M\rangle$ diverges in the degenerate case $\lambda=\mu$
if the mean waiting time $\langle T\rangle$ for catastrophic extinctions is infinite.

For $\lambda>\mu$, the integral in Eq.~(\ref{eB13})
establishes that unless $f(t)$ has at least exponential decay, the mean is necessarily divergent.
If
$f(t)\sim\hbox{constant}\times t^r\exp(-\theta t)$ as $t\to\infty$,
then the mean taxon size if finite so long as $\lambda<\mu+\theta$.  Whether it is also finite
in the critical case $\lambda=\mu+\theta$ depends on the value of $r$.  In particular,
for the exponential density 
$f(t)=\theta e^{-\theta t}$ we find that
$\langle M\rangle=\infty$ if $\lambda\geq\mu+\theta$, while
$\langle M\rangle=1+\lambda/(\theta+\mu-\lambda)$ if $\lambda<\mu+\theta$.

To analyse 
the case $\lambda>\mu$, we shall rewrite the integral for $\chi(\xi)$
in the equivalent form
\[
\chi(\xi)=\int_0^\infty {{\lambda\{1-e^{-(x_2-x_1)t}\}f(t)dt}
\over{(x_2-\lambda\xi) - (x_1-\lambda\xi)e^{-(x_2-x_1)t}}}.
\]
The exponentials are decaying functions of time, since 
$x_2-x_1=\lambda-\mu+2\lambda\mu(1-\xi)/(\lambda-\mu)+O([1-\xi]^2)$
as $1-\xi\to 0$.
Hence to leading order,
\[
\chi(\xi)\sim{\lambda\over{\lambda-\mu}}\int_0^\infty\kern-4pt {{[1-e^{-(\lambda-\mu)t}]f(t)dt}
\over{\kappa(1-\xi)
+e^{-(\lambda-\mu)t}}},\label{eB20}
\]
where $\kappa=\lambda^2/(\lambda-\mu)^2$.
In the case $f(t)=\theta e^{-\theta t}$, if we write $y=e^{-(\lambda-\mu)t}$,
we find that
\[
\chi(\xi)\sim{{\lambda\theta}\over{(\lambda-\mu)^2}}\int_0^1
{{(1-y)y^{\theta/(\lambda-\mu)-1}dy}
\over{\kappa(1-\xi)+y}}.
\]
Mellin transform methods
(see, e.g. \cite{bleistein} or \cite{hughes}, Appendix 2)
can be used to extract the asymptotic behavior of this integral and so determine the expansion for
$\phi(\xi)$ near $\xi=1$. We find that for $\lambda>\theta+\mu$,
\[
\phi(\xi)= 1-{{\pi\lambda[\kappa(1-\xi)]^{\theta/(\lambda-\mu)}}
\over{\theta\sin[(\pi\theta)/(\lambda-\mu)]}}+\cdots,
\]
while $\phi(\xi)=
1-(\lambda/\theta)(1-\xi)\ln [\kappa^{-1}(1-\xi)^{-1}]+\cdots$ for $\lambda=\theta+\mu$.
This asymptotic behavior of $\phi(\xi )$ suggests the following behavior of $p_m$ as $m\to\infty$:
\[
p_m\sim\left\{\begin{array}{ll}
\hbox{constant}\times m^{-1-\theta/(\lambda-\mu)},
&\lambda>\theta+\mu,\\
\hbox{constant}\times m^{-2}\ln  m&\lambda=\theta+\mu.
\end{array}
\right.
\]
To derive this rigorously would require either a careful argument based around 
Darboux's Theorem \cite{hughes}, or the methods of Flajolet and Odlyzko \cite{FlajoletandOdzlyko}, 
or Tauberian Theorems supplemented by information about the ultimate monotonic decay of $p_m$
\cite{hughes,feller}.  We obtain the same asymptotic behavior for the total number of 
species that ever existed as that found 
for currently living species:
\[
\Pr\{M\geq m\}\sim
\left\{\begin{array}{ll}
\hbox{constant}\times m^{-\theta/(\lambda-\mu)},& \lambda>\theta+\mu\\
\hbox{constant}\times m^{-1}\ln m, &\lambda=\theta+\mu.
\end{array}\right.
\]

More generally, note that as $\xi\to 1$,
\[
\chi(\xi)\sim{\lambda\over{\lambda-\mu}}\int_0^\infty\kern-4pt {{f(t)dt}
\over{\kappa(1-\xi)
+e^{-(\lambda-\mu)t}}}.
\]
It can be shown that if $f(t)\sim \Theta t^r e^{-\theta t}$ ($r>-1$)
then 
\[
\chi(\xi)\sim {{\pi\lambda\Theta[\kappa(1-\xi)]^{\theta/(\lambda-\mu)-1}
\{\ln [\kappa^{-1}(1-\xi)^{-1}]\}^r}
\over{(\lambda-\mu)^{r+2}\sin[(\pi\theta)/(\lambda-\mu)]}}
\]
for $\lambda>\theta+\mu$. In the borderline case $\lambda=\theta+\mu$, we find 
\[
\chi(\xi)\sim {{\lambda\Theta}\over{(r+1)\theta^{r+2}}}\,\{\ln [\kappa^{-1}(1-\xi)^{-1}]\}^{r+1}.
\]
We predict the asymptotic forms
\[
p_m\sim\left\{\begin{array}{ll}
\hbox{constant}\times m^{-1-\theta/(\lambda-\mu)}(\ln  m)^r,
&\lambda>\theta+\mu,\\
\hbox{constant}\times m^{-2}(\ln  m)^{r+1},
&\lambda=\theta+\mu.\end{array}\right.
\]

There has been a significant prior work on the modelling of populations subject to disasters in other
contexts,  with particular emphasis on the time to extinction in the process \cite{priorwork}. However, the
principal conclusions of the present paper, especially those drawn for the properties of extinct taxa,
appear to be new. We have shown that the competition between characteristic rates of 
species proliferation, individual species extinction, and large-scale catastrophic extinction
is able to generate long-tailed distributions of taxon size and consequent scaling properties and
fractal interpretations without the need to assume an underlying fractal model.  The formalism
covers both currently live taxa, and taxa destroyed out by previous global catastrophic 
extinction events.  Our results are based on a null
model for proliferations and extinctions.  The  validity of the model
can be assessed by comparing the results established with empirical size
distributions for living and fossil taxa (as in \cite{ReedHughesLiveTaxa}).
Models for evolution with an underlying dynamics
have been proposed \cite{null}. The null model provides a useful benchmark against which the
predictions of more detailed models may be assessed and its 
concepts and analytical methods may have applications in other areas.

\end{document}